\documentclass[conference]{IEEEtran}
\IEEEoverridecommandlockouts
\usepackage{fancyhdr}
\usepackage{cite}
\usepackage{amsmath,amssymb,amsfonts}
\usepackage{graphicx}
\usepackage{textcomp}
\usepackage{xcolor}
\usepackage{algorithm}
\usepackage{algpseudocode}
\usepackage{amsmath}
\def\BibTeX{{\rm B\kern-.05em{\sc i\kern-.025em b}\kern-.08em
    T\kern-.1667em\lower.7ex\hbox{E}\kern-.125emX}}
\begin{document}

\title{Backpressure-based Mean-field Type Game for Scheduling in Multi-Hop Wireless Sensor Networks\\
\thanks{This work was supported by the NYUAD Center for Interacting Urban Networks (CITIES), funded by Tamkeen under the NYUAD Research Institute Award CG001. The views expressed in this article are those of the authors and do not reflect the opinions of CITIES or their funding agencies.}
}

\author{\IEEEauthorblockN{1\textsuperscript{st} Salah Eddine Choutri}
\IEEEauthorblockA{\textit{NYUAD Research Institute} \\
\textit{New York University Abu Dhabi}\\
Abu Dhabi, UAE\\
choutri@nyu.edu}
\and
\IEEEauthorblockN{2\textsuperscript{nd} Boualem Djehiche}
\IEEEauthorblockA{\textit{Department of Mathematics} \\
\textit{KTH Royal Institute of Technology}\\
Stockholm, Sweden \\
boualem@kth.se}
\and
\IEEEauthorblockN{3\textsuperscript{rd} Prajwal Chauhan}
\IEEEauthorblockA{\textit{Engineering Division} \\
\textit{New York University Abu Dhabi}\\
Abu Dhabi, UAE\\
pc3377@nyu.edu}
\and
\IEEEauthorblockN{\quad\quad  4\textsuperscript{th} Saif Eddin Jabari}
\IEEEauthorblockA{\quad\quad \textit{Engineering Division} \\
\quad\quad\quad \ \textit{New York University Abu Dhabi}\\ \quad\quad\quad
Abu Dhabi, UAE\\ \quad\quad \ 
sej7@nyu.edu}
}
\maketitle 
\thispagestyle{fancy}

\begin{abstract}
We propose a Mean-Field Type Game (MFTG) framework for effective scheduling in multi-hop wireless sensor networks (WSNs) using backpressure as a performance criterion. Traditional backpressure algorithms leverage queue differentials to regulate data flow and maintain network stability. In this work, we extend the backpressure framework by incorporating a mean-field term into the cost functional, capturing the global behavior of the system alongside local dynamics. The resulting model utilizes the strengths of non-cooperative mean-field type games, enabling nodes to make decentralized decisions based on both individual queue states and system mean-field effects while accounting for stochastic network interactions. By leveraging the interplay between backpressure dynamics and mean-field coupling, the approach balances local optimization with global efficiency. Numerical simulations demonstrate the efficacy of the proposed method in handling congestion and scheduling in large-scale WSNs.
\end{abstract}


\section{Introduction}

Wireless Sensor Networks (WSNs) have emerged as a fundamental technology for large-scale monitoring applications, enabling data collection from a vast number of distributed sensor nodes \cite{akyildiz2002wireless, mainwaring2002wireless}. In multi-hop WSNs, data generated by sensor nodes must be efficiently relayed toward a designated sink node(s), requiring effective scheduling and congestion control strategies to optimize network performance. Traditional queue-based scheduling policies, such as the well-known Backpressure Algorithm \cite{tassiulas1992stability, neely2010stochastic}, provide solutions by prioritizing data transmissions based on queue differentials. Backpressure principles have been applied to a variety of real-world traﬃc problems, such as communication with autonomous vehicles, reservation systems, and vehicle routing \cite{zaidi2016back,li2019position,levin2016optimizing}.

However, solving backpressure-based optimization problems at large network scales is computationally demanding, especially due to 
the complexity of queue interactions. To mitigate these challenges, mean-field type games have emerged as a promising approximation framework that reduces the computational complexity of modeling large-scale queue interactions.

The concept of mean-field games was first introduced by Lasry and Lions \cite{lasry2007mean} and independently by Huang, Malhamé, and Caines \cite{huang2006large}, providing a framework for analyzing strategic interactions among a large number of agents influenced by aggregate system behavior. Stochastic Mean-field type control problems, also known as single-player mean-field type games, were, later, introduced by Andersson and Djehiche \cite{andersson2011}. These models typically describe stochastic optimization problems where the state dynamics and cost functional depend on the first moment of the state, introducing a mean-field coupling effect. The natural extension of this framework to multiple interacting agents leads to mean-field type games, which incorporate strategic decision-making among several players. This approach has been widely studied in various contexts (see e.g. \cite{agram2020mean, choutri2019mean}).

Building on this foundation, we propose a Mean-Field Type Game (MFTG) framework that extends backpressure-based scheduling by incorporating mean-field interactions for scalable decision-making in multi-hop WSNs. While traditional backpressure algorithms regulate data flow based on local queue differentials, they do not account for large-scale network effects. By integrating a mean-field term into the decision process, our approach enables decentralized scheduling where nodes optimize transmission strategies based on both local queue states and system congestion dynamics, improving scalability and efficiency.

To address the complex queue interactions, we heuristically derive a mean-field type approximation for the queueing dynamics, justifying its validity through key assumptions about network exchangeability and weak interactions \cite{oelschlager1984martingale}. This allows us to replace empirical averages with expectations, which reduces computational complexity.

The main contributions of this paper are as follows:
\begin{itemize}
    \item We propose a game-theoretic model for transmission scheduling in multi-hop WSNs, extending the classical backpressure framework to a non-cooperative mean-field type setting.
    \item We heuristically derive the mean-field type queue dynamics and justify the approximation based on network assumptions.
    A rigorous mathematical derivation of such a mean-field type dynamics based on the law of large numbers is outside the scope of this paper. It will appear elsewhere.
    \item We validate our approach through numerical simulations, demonstrating its effectiveness in balancing the load among the nodes. 
\end{itemize}

The remainder of this paper is structured as follows: Section II introduces the network model and describes the queue dynamics of multi-hop WSNs. Moreover, it presents the backpressure-based scheduling game framework, while Section III derives the mean-field type game. Section IV presents simulation results validating the proposed approach, and finally, Section V concludes with key insights and future research directions.

\section{System Model}
\subsection{Network Model}
We consider a multi-hop wireless sensor network for continuous data flow, where multiple sensor nodes generate data that must be transmitted to a single sink node (or multiple sinks). The network is modeled as a directed graph \( G = (\mathcal{N}, E) \), where

\begin{itemize}
    \item \( \mathcal{N}\) is the set of nodes, consisting of \( N \) sensor nodes and one designated sink node \( d \).
    \item \( E \) is the set of directed wireless links, where \( (i, j) \in E \) indicates that node \( i \) can transmit data to node \( j \).
    \item Each sensor node \( i \in \mathcal{N}, i \neq d \) generates a continuous flow of data that must be delivered to the sink node \( d \).
\end{itemize}

Each sensor node \( i \in \mathcal{N}, i \neq d \) maintains a queue \( Q_i(t) \), representing the amount of buffered data (measured in a suitable data units, e.g. bytes). Nodes (sensors) make transmission decisions using an on-off scheduling policy, where the control variable \( \chi_i \) determines whether a node transmits data or remains idle. Additionally, sensor nodes may (partially) process some incoming or locally generated data before forwarding it, reducing the queue length through data compression, aggregation, or event-driven filtering. This is particularly relevant in applications such as environmental monitoring, where redundant sensor readings can be combined into a single representative value, or industrial IoT, where periodic machine status updates can be aggregated. \\
Each queue is modeled as a \( \mathbb{R}_+ \)-valued stochastic queuing process, \( Q_i = \{Q_i(t), t \geq 0\} \), where at each time $t$, the queue length is described by the following equation:

\begin{equation} \label{queue}
     Q_i(t) = Q_i(0)+ A_i(t) + F_i(t) - D_i(t) - P_i(t),
\end{equation}
where:
\begin{itemize}
    \item $Q_i(0)$ is assumed $0$ without loss of generality.
    \item \( A_i(t) \in \mathbb{Z}_+ \) is the external data arrivals at node \( i \). 
    \item \( D_i(t) \in \mathbb{Z}_+ \) is the total outgoing transmissions from node \( i \) that depend on the network control status $\chi_i \in \{0,1\}$.
    \item \( F_i(t) \in \mathbb{R}_+ \) is the total incoming data received from upstream nodes.
    \item \( P_i(t) \in \mathbb{R}_+ \) is the total data processed (e.g. aggregated).
\end{itemize}

The "aggregator" process at any time $t$, \(P_i(t)\), for any node $i$ is modeled as
\begin{equation} \label{aggregator}
    P_i(t) = \beta (A_i(t)+F_i(t)), \quad \beta \in [0,1].
\end{equation} where $\beta$ is the aggregation factor.\\
Replacing \eqref{aggregator} in \eqref{queue}, the queue length's equation becomes
\begin{equation}
    Q_i(t) = (1-\beta) [A_i(t) + F_i(t)] - D_i(t), \forall i \in \mathcal{N} \setminus \{d\}.
\end{equation} \\
The sink node \( d \) continuously removes all received data, ensuring
\begin{equation}
    Q_d(t) = 0, \quad t\ge 0.
\end{equation}
\\
The data arrival process at node \( i \) follows the process
\begin{equation}
    A_i(t) = N_i^+ \left( \int_0^t \lambda_i(u) \, du \right),
\end{equation}
where
\begin{itemize}
    \item \( \lambda_i : \mathbb{R}_+ \to \mathbb{R}_+ \) are the data arrival rates, we assume that they are randomized and i.i.d.
    \item \( N_i^+(\cdot) \) is rate 1 right-continuous with left-limits (RCLL) stochastic point processes (Poisson counting process), meaning each event corresponds to a burst of incoming data.  
\end{itemize}

The departure rate of node \( i \) is given by
\begin{equation}
    D_i(t) = N_i^- \left( \int_0^t \mu_i(Q_i(u)) \chi_i(u) \, du \right),
\end{equation}
where
\begin{itemize}
    \item \( N_i^-(\cdot) \) is a rate 1 Poisson process independent of \( N_i^+(\cdot) \). Each Poisson departure event corresponds to a transmission of a burst of data as well.
    \item We model $\mu_i$ as $\mu_i(Q_i(u))= \frac{m_i}{1+\alpha Q_i(u)}, \forall i,$ where $\alpha \in [0,1]$ and \( m_i : \mathbb{R}_+ \to [0, m_{\text{max}}] \) are measurable, base transmission rate, functions with bounded support (\( m_{\text{max}} \) is a finite fixed scalar). We assume that these rates $(m_i's)$ are random and $i.i.d.$
\end{itemize}
This formulation ensures that nodes can adaptively regulate their transmission rates based on their queue occupancy and network conditions. \\

The total \textit{incoming data flow rate} at node \( i \) is given by
\begin{equation}
    F_i(t) = \sum_{j \in \mathcal{N}_i} \int_0^t \frac{1}{\vert \mathcal{N}_i \vert} dD_j(u),
\end{equation}
where $\mathcal{N}_i$ is the set of neighboring nodes of $i$ and $\vert \mathcal{N}_i \vert$ denotes its cardinality. Note that we assume a fixed uniform routing, i.e, $\sum_{j \in \mathcal{N}_i} \frac{1}{\vert \mathcal{N}_i \vert}=1, \ \forall i.$

The expected cumulative arrival process satisfies
    $
        \mathbb{E}[A_i(t)] = \int_0^t \lambda_i(u) \, du,
    $
    while the expected cumulative departure process is given by
    $
        \mathbb{E}[D_i(t)] = \int_0^t \mu_i(Q_i(u)) \chi_i(u) \, du.
    $

\noindent The scaled arrival processes and the departure process have the following associated martingales:
\begin{itemize}
    \item $M^A_i(t)= (1-\beta)[A_i(t)-\int_0^t \lambda_i(u) \, du]$.
    \item $M^D_i(t)= D_i(t)- \int_0^t \mu_i(Q_i(u)) \chi_i(u) \, du $.
    \item
    \begin{align*}
        M^F_i(t) &=  
        \begin{aligned}[t]
            &(1-\beta) \Big[ F_i(t) \\
            &\quad - \int_0^t \sum_{j \in N_i} \frac{1}{\vert \mathcal{N}_i \vert} \, \mu_j(Q_j(u)) \chi_j(u) \, du \Big].
        \end{aligned}
    \end{align*}

\end{itemize}
The compensator for the martingale $M_i^F(t)$ is computed from the following.
\begin{align*}
    (1-\beta)\mathbb{E}[F_i(t)] &= (1-\beta) \sum_{j \in N} \mathbb{E}\left[\int_0^t \frac{1}{\vert \mathcal{N}_i \vert} \, dD_j(u)\right]\\
                       &= (1-\beta) \sum_{j \in N} \int_0^t \frac{1}{\vert \mathcal{N}_i \vert} \, \mu_j(Q_j(u)) \chi_j(u) \, du \ \\ &\text{(since $\mathbb{E}[dD_j(u)] = \mu_j(Q_j(u)) \chi_j(u) \, du)$} \\
                       &=(1-\beta)\int_0^t \sum_{j \in N_i} \frac{1}{\vert \mathcal{N}_i \vert} \, \mu_j(Q_j(u)) \chi_j(u) \, du.
\end{align*}

\vspace{0.5cm}
\noindent We rewrite $Q_i$ as

\begin{align*}
    Q_i(t) =& M_i^A(t) - M_i^D(t) + M_i^F(t)\\
    &+ (1-\beta) \int_0^t \lambda_i(u) \, du \
    \ - \int_0^t \mu_i(Q_i(u)) \chi_i(u) \, du  \\
    &+ (1-\beta) \int_0^t \sum_{j \in N_i} \frac{\mu_j(Q_j(u)) \chi_j(u)}{\vert \mathcal{N}_i \vert} \, du \notag \\
    =& M_i^Q(t) +  \int_0^t (1-\beta) \Bigg(\lambda_i(u) + \sum_{j \in N_i} \frac{\mu_j(Q_j(u)) \chi_j(u)}{\vert \mathcal{N}_i \vert} \Bigg)  \\ 
    &\quad\quad\quad\quad\quad\quad\quad\quad\quad\quad\quad\quad\quad
    - \mu_i(Q_i(u)) \chi_i(u) \ du. \notag
\end{align*}

\noindent where $M_i^Q$ is the martingale associated with \( Q_i \). \\

The differential form is given by

\begin{align} \label{dynamics}
    dQ_i(t) &= \Bigg((1-\beta)\lambda_i(t) + \sum_{j \in N_i} \frac{(1-\beta)}{\vert \mathcal{N}_i \vert} \mu_j(Q_j(t)) \chi_j(t)  \\
    &\quad - \mu_i(Q_i(t)) \chi_i(t) \Bigg) dt + dM_i^Q(t). \notag
\end{align}
It is worth noting that although Poisson processes count discrete transmission events, the queue length is measured in some data unit rather than packets. Since the Poisson intensity is integrated over time, the queue evolution smooths out in expectation, justifying a continuous-time differential equation.
\subsection{Backpressure-Based Scheduling Game Framework}
The traditional formulation of backpressure-based scheduling problems can be seen a cooperative game that consists of fixing a time instance \( t \), at which all the nodes (players) determine collectively the optimal transmission policies \( \chi^* \) that maximize the over all backpressure-based criterion, i.e.
\begin{equation}
    \chi^*(t) \in \arg\max_{\chi \in I(G)} \ \mathbb{E}\sum_{i=1}^{N} \sum_{j \in \mathcal{N}_i}  \frac{1}{\vert \mathcal{N}_i \vert} \big( Q_i(t) - Q_j(t) \big) \mu_i(Q_i(t)) \chi_i,
\end{equation}
Subject to \eqref{dynamics},\ 
with
\begin{itemize}
    \item \( \chi_i \in \{0,1\} \) determines wether the node is enabled for transmission.
    \item \( I(G) = \{0,1\}^N \) represents the set of feasible transmission schedules.
    \item The term \( (Q_i - Q_j) \) ensures that data moves toward less congested nodes, following backpressure principles.
\end{itemize}

The non-cooperative formulation models each node as a self-interested player making decisions that are locally optimal based on its own queue state and transmission opportunities. Unlike cooperative strategies, where nodes coordinate to achieve a shared objective (optimize a social utility), each node in this setting acts selfishly to maximize its own transmission efficiency. However, nodes are still indirectly influenced by the states of their neighboring queues, as seen in the utility function \eqref{utility1}, where the transmission decision depends on the relative queue differentials \(Q_i(t) - Q_j(t)\). This structure naturally captures competition for transmission opportunities and the impact of congestion.
\begin{equation} \label{utility1}
    U_i(t, \chi_i, \chi_{-i}) = \mathbb{E} \left[ \sum_{j \in \mathcal{N}_i}  \frac{1}{\vert \mathcal{N}_i \vert} \big( Q_i(t) - Q_j(t) \big) \mu_i(Q(t)) \chi_i \right],
\end{equation}
where \( \chi_{-i} \) represents the actions of all other players in the neighborhood of \( i \) that corresponds to the queues $Q_j, \ j \in \mathcal{N}_i$.

The best response for node \( i \) at time \( t \) is given by
\[
\max_{\chi \in I(G)} U_i(t, \chi_i, \chi_{-i}),
\]
subject to \eqref{dynamics}. 

\section{Mean-field type game formulation}




As the number of neighboring nodes grows large, the influence of any single neighbor is negligible. In the limit, we obtain the mean-field queue dynamics
\begin{align}
    dQ_i(t) =& \Bigg((1-\beta)\lambda_i - \mu_i(Q_i) \chi_i + (1-\beta)\mathbb{E}[\mu_i(Q_i) \chi_i] \Bigg) dt \notag\\
    &\quad\quad + dM_i^Q(t).
\end{align}
This equation maintains node-level heterogeneity while incorporating a mean-field correction term that captures interactions at scale. To justify replacing empirical averages with expectations, we first justify the exchangeability of nodes. Exchangeability means that nodes are statistically identical, meaning that their joint distribution is invariant under permutations. This assumption is valid in large-scale sensor networks under the following conditions: 
\begin{itemize}
    \item Homogeneous sensor deployment where sensors are deployed densely and uniformly, meaning that their arrival rates and base service rates are independent and follow the same statistical distribution,
    \item Dense local interactions ensuring that any single node’s effect on another is diluted as the network grows,
    \item Identical decision rules where each node solves the same optimization problem and follows an identical decision-making process, meaning that the control policy \( \chi_i \) is drawn from the same functional form across all nodes,
    \item Network graph regularity, where the underlying network topology is sufficiently regular, meaning that all nodes experience similar levels of congestion and interference.
\end{itemize}
Under these conditions, the empirical mean of neighboring queues converges to its expectation by the law of large numbers:
\begin{equation}
\frac{1}{|\mathcal{N}_i|} \sum_{j \in \mathcal{N}_i} Q_j(t) \approx \mathbb{E}[Q_i(t)] \quad \text{as } |\mathcal{N}_i| \to \infty.
\end{equation}
This approximation allows us to express queue interactions in a mean-field form. Furthermore, under the assumptions of weakly interacting queues implying exchangeability, \textit{propagation of chaos} ensures that individual nodes become decoupled, effectively reducing direct pairwise dependencies. This decoupling significantly lowers computational complexity, as each node’s decision process depends only on its local state and a mean-field term rather than explicit interactions with neighbors.
Thus, the utility function in \eqref{utility1} simplifies to
\begin{equation}
    U_i(t, \chi_i^*, \chi_{-i}) = \max_{\chi_i \in I(G)} \mathbb{E} \left[ (Q_i(t) - \mathbb{E}[Q_i(t)]) \mu_i(Q_i) \chi_i \right].
\end{equation}
This equation shows that each queue evolves based on its own arrival and departure process, a mean-field term that captures the impact of all other nodes in the network, and the control \( \chi_i \) chosen based on local and mean-field effects.


\section{Simulations}
\begin{figure}[htbp]
    \centering
    \includegraphics[width=\columnwidth, height=0.2\textheight, keepaspectratio]{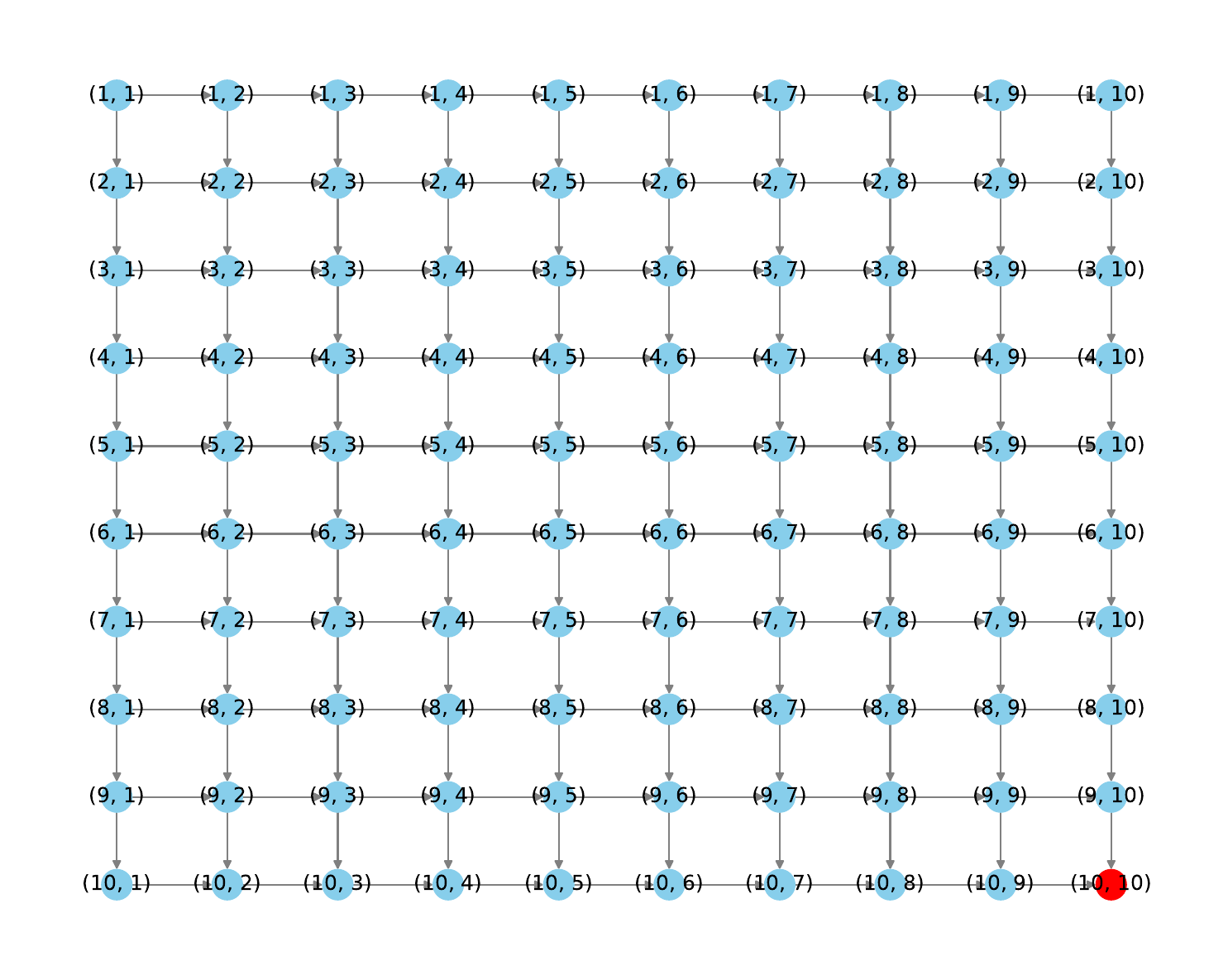}
    \caption{Directed Grid with last node as sink}
    \label{fig:grid}
\end{figure}

In this section we simulate a dense network modeled as a directed acyclic graph. The network comprises 250000 nodes (including one sink node) arranged in a grid configuration, (see Figure~\ref{fig:grid}). Each node, here, represents a homogeneous group of sensors that are behaving similarly. Moreover,
we consider a discrete-time approach to simulate the (mean-field) evolution of the queuing process \( Q_i(t) \) for each node \( i \) in the limit, which is governed by a combination of external arrivals, internal routing from other nodes, and departures due to transmissions and "partial" internal processing. 
Let \( [0, T] \) be a fixed time interval that is divided into uniform time steps \( t_k = k \cdot \Delta t \), where \( \Delta t \) is the discrete time step, and \( k \in \{0, 1, \dots, K\} \), such that \( K = \frac{T}{\Delta t} \). The discretized queue dynamics is simulated, for each node $i$, as
\begin{align*}
Q_i(t_k+ \Delta t) = Q_i(t_{k}) &+(1-\beta) [A_i(t_{k} + \Delta t ) +  F_i(t_{k} + \Delta t)] \\ &- D_i(t_{k} + \Delta t).
\end{align*}

At each time step \( t_k \), the number of external arrivals to node \( i \) is sampled from a Poisson distribution with rate \( \lambda_i(t_k) \) and scaled by the factor $(1-\beta)$, while the number of departures is also sampled from a Poisson distribution but with  rate \( \mu_i(Q_i(t_k)) \) \( \chi_i(t_k) \), i.e., $A_i(t_{k+1}) \sim \text{Poisson}(\lambda_i(t_k) \Delta t), \  D_i(t_{k+1}) \sim \text{Poisson}(\mu_i(Q_i(t_k)) \chi_i(t_k) \Delta t),$ where the the arrival rates $\lambda_i's$ are sampled independently from a uniform distribution.

We simulate the internal arrivals for each node $i$ by computing, empirically, the scaled quantity
 $(1-\beta) \mu_i(Q_i) \chi_i] \Delta t $. For simplicity, we adopt a normalized time step of $\Delta t=1$. The transmission rates $\mu_i(Q_i)$ are modeled as $m_i / \left(1 + \alpha Q_i(t)\right), \ \alpha>0$, where the base rates $m_i's$ are independent and uniformly distributed. At each time instance, we determine the value of $\chi_i,$ for each node, depending on the value of the differential  $(Q_i(t) - \mathbb{E}[Q_i(t)]).$ The control $\chi_i$ is one if the differential is positive and zero otherwise. Table \ref{parameters} summarizes the values of all the parameters used in the simulation, while Algorithm \ref{Algorithm} details the simulation. 

\begin{figure}[htbp]
    \centering
    \includegraphics[width=\columnwidth, height=0.5\textheight, keepaspectratio]{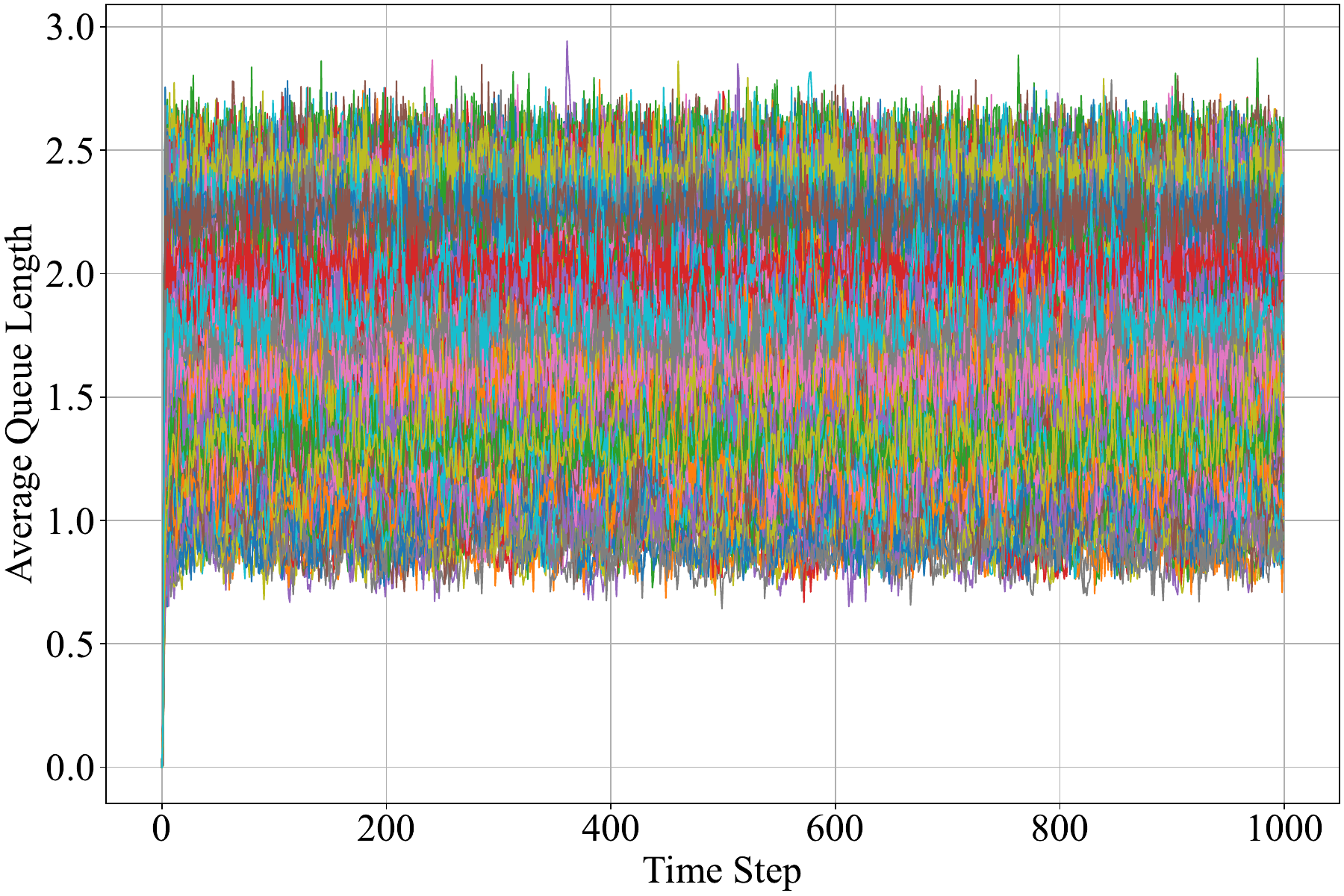}
    \caption{A simulated system of 250,000 nodes (1000 are shown)}
    \label{fig:queue_simulation2}
\end{figure}

Figure \ref{fig:queue_simulation2} illustrates the behavior of average queue lengths. Initially, we observe a rapid increase in the queue lengths, which is due to the difference between data arrivals and the limited initial transmission capacity. As the system evolves, the accumulation of data gradually builds sufficient backpressure, which helps to regulate the flow of data, resulting in a more balanced state where the queue lengths stabilize at a consistent level. Such a transition, from an initial phase of instability to a steady state, is characteristic of systems employing backpressure, where the algorithm requires some time to adjust and optimize the data flow before achieving equilibrium.

\begin{algorithm}[ht]
\caption{Backpressure-based Mean-field Type Pseudocode}
\label{Algorithm} 
\begin{algorithmic}[1] \footnotesize
\Require Parameters: $\beta$, $m_{\min}$, $m_{\max}$, $\lambda_{\min}$, $\lambda_{\max}$, $K$, $N$, $M$, $\alpha$
\rule{\linewidth}{0.4pt}
\State \textbf{Initialization:} For each node $i=1,\dots,N$, and for each sample $j=1,\dots,M$:
\State \quad Set $Q_i^j(0)=0$ and $\chi_i(0)=1$.
\State \quad Sample $\lambda_i\sim\mathcal{U}(\lambda_{\min},\lambda_{\max})$ and $m_i\sim\mathcal{U}(m_{\min},m_{\max})$.
\For{$k=1,\dots,K$}
    \For{$i=1,\dots,N$}
        \For{$j=1,\dots,M$}
            \State Generate $A_i^j(k)\sim\text{Poisson}(\lambda_i)$.
            \State Compute $\mu_i^j(k)=\dfrac{m_i}{1+\alpha\,Q_i^j(k-1)}$.
            \State Generate $D_i^j(k)\sim\text{Poisson}\Bigl(\mu_i^j(k)\,\chi_i(k-1)\Bigr)$.
            \State Compute the internal arrival rate:
            $F_i^j(k)=\mu_i^j(k)\,\chi_i(k-1).$
            \State Update the queues:
            \[
            Q_i^j(k)=Q_i^j(k-1)+(1-\beta)\Bigl[A_i^j(k)+F_i^j(k)\Bigr]-D_i^j(k).
            \]
        \EndFor
        \State Set $Q_{\text{sink}}(k)=0$.
        \State Compute the sample average:
        $\bar{Q}_i(k)=\frac{1}{M}\sum_{j=1}^{M}Q_i^j(k).$
        \State Update control:
        \[
        \chi_i(k)=
        \begin{cases}
          1, & \text{if } Q_i^r(k)>\bar{Q}_i(k), \quad r \sim \mathcal{U}\{1, \dots, M\}, \\[1mm]
          0, & \text{otherwise.}
        \end{cases}
        \]
    \EndFor
\EndFor
\end{algorithmic}
\end{algorithm}

\begin{table}[ht]
    \centering
    \caption{Simulation Parameters}
    \begin{tabular}{cll} \label{parameters}
        \textbf{Parameter} & \textbf{Description} & \textbf{Value/Range} \\
        \hline
        $N$ & Number of queues\slash nodes & 250000 \\
        $K$ & Total simulation time & 1000 time units \\
        $\Delta t$ & Time step size  & 1 time units \\
        $M$ & number of "trial" samples & 100 \\
        $m_i$ & Base service rate & $\mathcal{U}[1, 5]$ \\
        $\alpha$ & Congestion sensitivity factor & 0.01 \\
        $\beta$ & Aggregation factor & 0.7 \\
        $\lambda_i$ & External arrival rate & $\mathcal{U}[0.1,0.5]$ \\
        $Q(0)$ & Initial queue length  & $0$ \\
        \hline
    \end{tabular}
    \label{tab:simulation_parameters}
\end{table}
\color{black}


\section*{Conclusion}

This work introduced a Mean-Field Type Game formulation for decentralized scheduling in multi-hop wireless sensor networks using backpressure-inspired principles. By incorporating a mean-field interaction term into the node dynamics, the model enables local decision-making while approximating large-scale network behavior.
We modeled the system as a non-cooperative game, where each node optimizes its transmission strategy based on local queue states and the average behavior across the network. Through a law of large numbers approximation, we derived a stochastic differential equation that captures the limiting dynamics of the system while reducing the complexity associated with tracking individual queue interactions.
While our numerical simulations illustrate the consistency of the mean-field approximation and show stable behavior under varying node counts, a comprehensive performance comparison with classical backpressure scheduling and other control strategies is beyond the current scope.
Future work includes quantitatively benchmarking our approach against conventional backpressure methods, studying heterogeneous topologies, adaptive learning strategies, and interference-aware mechanisms to enhance the model’s applicability to real-world wireless networks. In addition, we plan to analyze the impact of interference constraints in the mean-field regime and demonstrate how the original (pairwise) constraints transition into a statistical activation constraint.

\bibliographystyle{IEEEtran}  

\end{document}